\documentclass[twocolumn, superscriptaddress,pra,aps,showpacs]{revtex4-1}
\usepackage{amsmath,mathrsfs,amsbsy,color,graphicx,bm,amsthm,amsfonts,array,multirow}
\usepackage[english]{babel}
\usepackage[latin1]{inputenc}
\usepackage{graphicx}
\usepackage{braket}
\usepackage{tabularx}
\usepackage{afterpage}

\newtheorem{principle}{Principle}
\newtheorem{observation}[principle]{Observation}

\newtheorem{theorem}[principle]{Theorem}

\begin{document}

\title{Multipartite states under elementary local operations}

\author{F. E. S. Steinhoff} 
\email{steinhofffrank@gmail.com}
\affiliation{Faculdade de Engenharia, Universidade Federal de Mato Grosso, 78060-900 V\'arzea Grande, Mato Grosso, Brazil}


\date{\today}

\begin{abstract} 
Multipartite pure states are equivalent under Stochastic Local Operations and Classical Communication (SLOCC) whenever they can be mapped into one another by Invertible Local Operations. It is shown that this is equivalent to the inter-convertibility through finite sequences of Elementary Local Operations. A multipartite version of the Gauss-Jordan elimination strategy is then obtained, enabling the analytical SLOCC-classification of previously unknown examples. It is argued that the problem of SLOCC-classification is equivalent to the problem of classifying the Multipartite Fully Reduced Forms that the coefficient matrix of a state can assume after being subjected to this Gauss-Jordan elimination procedure. The method is applied to examples of so-called LME and hypergraph states, showing new SLOCC-equivalences. Moreover, possible physical implications for states with a Dicke-like structure are sketched.       
\end{abstract}

\maketitle

\section{Introduction}

The classification of multipartite states with respect to Stochastic Local Operations and Classical Communication (SLOCC) is a central problem in Quantum Information Theory and a very challenging one. It is mostly related to the problem of finding sets of genuinely multipartite entangled states and their uses in quantum computing and information protocols as a resource \cite{nielsen}, but its importance is not restricted to entanglement theory. It ranges from applications such as metrology \cite{metrology}, resource theories \cite{resourcetheories}, magic-state distillation \cite{magicstates} and many-body physics \cite{spto}, to more fundamental aspects such as nonlocality \cite{nonlocality}, the marginal problem \cite{margprob}, nonclassicality \cite{nonclassicality} and quantum thermodynamics \cite{quantthermo}.

Since pioneering \cite{wghz} studies, many ground-breaking results on the subject were developed. In \cite{miyake}, a classification in terms of hyperdeterminants was obtained; in \cite{verstraete}, large sets of states were shown to have a normal form characteristic of their SLOCC class. More recently, \cite{lamata,bkraus} have shown criteria in great generality. For a review on the subject, one can check \cite{multientang,gtreview}; the reference \cite{cris} presents a more recent overview.

Inspired by these and many other results on the subject, the current work presents a method of determining the SLOCC-equivalence of states by the use of Elementary Local Operations. These are tensor products of well-known Elementary Operations \cite{linalgebra} in linear algebra, used, among other applications, in the procedure known as Gauss-Jordan Elimination, which characterizes systems of linear equations. Elementary Local Operations are the building blocks of Invertible Local Operations and thus of SLOCC manipulations in a multipartite pure state. Physically, Invertible and Elementary Local Operations correspond to processes of local filtering \cite{localfilters}. As in its standard uses, Gauss-Jordan elimination in its multipartite version is shown to be a very powerful procedure, enabling by-hand derivations of previously complicated results. Moreover, it is argued that the SLOCC-classification problem is equivalent to the classification of the possible Multipartite Fully Reduced Forms of a system, the analogues of Echelon Forms in standard linear algebra. 

The article is divided as follows: Section II gives an overview of important concepts such as coefficients matrices, the SLOCC-classification problem for multipartite pure states and Elementary Operations and Matrices. Section III shows (a) that two multipartite pure states are SLOCC-equivalent if and only if (iff) they can be mapped into each other by sequences of Elementary Local Operations; (b) that an Elementary Local Operation induces a collective Elementary Operation on the coefficient matrix of a state; (c) the multipartite analogue of the Gauss-Jordan elimination procedure; and (d) two illustrative simple examples. In Section IV, the method is applied to the families of genuinely entangled multipartite states known as Locally Maximally Entangleable (LME) states \cite{lme} and hypergraph states \cite{hypergraphpapers,hyper1,hyper2}, showing analytically new SLOCC-equivalences. Section V presents remarks on potential physical aspects related to the SLOCC-equivalence of superpositions of Dicke-like states. Section VI contains conclusions and perspectives and VII the acknowledgments. 



\section{Preliminaries}

Let $|\psi\rangle\in\mathcal{H}$, where $\mathcal{H}=\bigotimes_{k=1}^N\mathcal{H}_k$ is a $N$-partite system and the subsystems $\mathcal{H}_k$ have each dimensions $d_k$. Let $\{|i_1i_2\ldots i_N\rangle\}$ denote the computational basis of $\mathcal{H}$, with $i_k=0,1,\ldots,d_k-1$. Then $|\psi\rangle$ can be expressed as 
\begin{eqnarray}
|\psi\rangle = \sum_{i_k=1}^{d_k-1} c_{i_1i_2\ldots i_N}|i_1i_2\ldots i_N\rangle \label{expansion}
\end{eqnarray}
The scalars $c_{i_1i_2\ldots i_N}$ can be arranged in a matrix $C_{|\psi\rangle}$, the coefficient matrix of the state $|\psi\rangle$, with rows indexed by the values $i_1=0,1,\ldots,d_1-1$ and columns arranged according to the other $i_k$'s. The simplest example is that of two qubits, where an arbitrary state is written as
\begin{eqnarray*}
|\psi\rangle = c_{00}|00\rangle +c_{01}|01\rangle + c_{10}|10\rangle+c_{11}|11\rangle   
\end{eqnarray*}
and its coefficient matrix reads
\begin{eqnarray}
C_{|\psi\rangle} = \left(\begin{array}{cc}
						{c_{00}}&{c_{01}} \\
						{c_{10}}&{c_{11}}
\end{array}\right).
\end{eqnarray} 
For the three-qubit case, an arbitrary state is given by
\begin{eqnarray*}
|\psi\rangle &=& c_{000}|000\rangle +c_{001}|001\rangle + c_{010}|010\rangle + c_{011}|011\rangle \\
             &=& c_{100}|100\rangle +c_{101}|101\rangle + c_{110}|110\rangle + c_{111}|111\rangle,  
\end{eqnarray*}
and its coefficient matrix is 
\begin{eqnarray}
C_{|\psi\rangle} = \left(\begin{array}{cc|cc}
						{c_{000}}&{c_{001}}&{c_{010}}&{c_{011}} \\
						{c_{100}}&{c_{101}}&{c_{110}}&{c_{111}}
\end{array}\right).
\end{eqnarray} 
For $N$ qudits ($d_1=d_2=\ldots=d_N=d$), the coefficient matrix of a state has then the format
\begin{eqnarray}
C_{|\psi\rangle} = \left(\begin{array}{c|c|c|c}
						{M_0}&{M_1}&{\ldots}&{M_{d^{N-2}-1}} 
					\end{array}\right),
\end{eqnarray} 
where each $M_k$ is a $d\times d$ block. In this work, the focus mostly rely on examples where all subsystems have equal dimension $d$, but the formulation can be easily extended to more general situations. 

Moreover, we say that $|\phi\rangle$ and $|\psi\rangle$ are SLOCC-equivalent if one can be obtained from the other by the sole use of SLOCC procedures. Even though the classification of SLOCC maps is in itself a complicated task, for pure states the problem simplifies considerably due to the following observation \cite{wghz}:  
\begin{observation}
Two pure $N$-partite states $|\phi\rangle$ and $|\psi\rangle$ are SLOCC-equivalent iff there exist Invertible Local Operators (ILOs) $A_k$ such that
\begin{eqnarray*}
|\phi\rangle = A_1\otimes A_2\otimes\ldots\otimes A_N|\psi\rangle.
\end{eqnarray*}
\end{observation} 
An important special case is when one imposes the extra requirement that the $A_k$'s comprise unitary operators. If this stronger constraint is satisfied, we say that $|\phi\rangle$ and $|\psi\rangle$ are equivalent under Local Unitary Operators, or LU-equivalent. 

\subsection{Elementary Operations}

An elementary row-operation on a matrix consists of one of the following operations:
\begin{enumerate}
\item $F(i,j)$: Permutation of rows $i$ and $j$.
\item $S(k,\lambda)$: Multiplication of row $k$ by a non-zero scalar $\lambda$.
\item $L(i,\lambda,j)$: Sum of row $i$ times $\lambda$ to row $j$. 
\end{enumerate}
Likewise, one can define elementary column-operations on a matrix in a similar fashion. Whenever necessary, we use only the term Elementary Operation and the symbol $E$ for a generic operation of this kind.  

Each elementary row-operation on a $m\times n$ matrix $A$ corresponds to the left-multiplication of $A$ by an $m\times m$ Elementary Matrix. Operation $F(i,j)$ on $A$ is the same as performing $F(i,j)A$, where
\begin{eqnarray}
F(i,j) = |i\rangle\langle j| + |j\rangle\langle i| + \sum_{k\neq i,j} |k\rangle\langle k|
\end{eqnarray}
while operations $S(k,\lambda)$ and $L(i,\lambda,j)$ on $A$ correspond respectively to $S(k,\lambda)A$ and $L(i,\lambda,j)A$, with
\begin{eqnarray}
S(k,\lambda) &=& \lambda|k\rangle\langle k| +\sum_{j\neq k}|j\rangle\langle j| \\
L(i,\lambda,j) &=& \lambda|j\rangle\langle i| + I_m 
\end{eqnarray}
where $I_m$ denotes the $m\times m$ identity matrix. Conversely, elementary column-operations on a $m\times n$ matrix $A$ are implemented by right-multiplication of $A$ by an appropriate $n\times n$ Elementary Matrix. For example, the set of elementary $2\times 2$ matrices is given by
\begin{eqnarray*} 
F(0,1)=F(1,0)=\left(\begin{array}{cc}
{0}&{1}\\
{1}&{0}
\end{array}\right); \\ 
S(0,\lambda)=\left(\begin{array}{cc}
{\lambda}&{0}\\
{0}&{1}
\end{array}\right); \ \ 
S(1,\lambda)=\left(\begin{array}{cc}
{1}&{0}\\
{0}&{\lambda}
\end{array}\right); \\ 
L(0,\lambda,1)=\left(\begin{array}{cc}
{1}&{0}\\
{\lambda}&{1}
\end{array}\right); \ \ 
L(1,\lambda,0)=\left(\begin{array}{cc}
{1}&{\lambda}\\
{0}&{1}
\end{array}\right).  
\end{eqnarray*}
Sequences of Elementary Operations are the key ingredient in the method known as Gauss-Jordan elimination (GJE), the main strategy to characterize the solutions of systems of linear equations and reviewed briefly in Appendix A. 

Moreover, elementary matrices are fundamental due to the following fact \cite{linalgebra}:
\begin{theorem}
Every invertible operation is the composition of a finite sequence of elementary operations.
\end{theorem}
Indeed, elementary matrices and operations are the generators of the so-called general linear group $GL(n)$ of $n\times n$ invertible matrices. This means that every invertible matrix is a product of elementary matrices. To find the elementary matrix that implements a given elementary operation on a matrix, one applies this operation to the identity operator. Thus a product of elementary matrices is an invertible operator, since applying the inverse sequence of operations (in reverse order) brings this operator into the identity again. For example:
\begin{eqnarray*}
M&=&\left(\begin{array}{ccc}
{1}&{0}&{0}\\
{2}&{1}&{0}\\
{0}&{4}&{5}
\end{array}\right)=L(0,2,1)L(1,4,2)S(2,5). \\
M^{-1}&=&\left(\begin{array}{ccc}
{1}&{0}&{0}\\
{-2}&{1}&{0}\\
{8/5}&{-4/5}&{1/5}
\end{array}\right)\\ &=& S(2,1/5)L(1,-4,2)L(0,-2,1).
\end{eqnarray*}       

\section{General considerations}
Let $E_i^{(l_i)}$ denote a sequence of compositions of Elementary Local Operations (ELOs) on the subsystem $\mathcal{H}_i$. Combining Theorem 2 and Observation 1, we arrive at the following statement:
\begin{theorem}
Two pure $N$-partite states $|\phi\rangle$ and 
$|\psi\rangle$ are SLOCC-equivalent iff there exists a sequence of ELOs $E^{(l_1)}_1,E^{(l_2)}_2,\ldots,E^{(l_N)}_N$ such that
\begin{eqnarray}
|\phi\rangle = E^{(l_1)}_1\otimes E^{(l_2)}_2\otimes\ldots\otimes E^{(l_N)}_N|\psi\rangle.
\end{eqnarray}
In this case, we say that $|\phi\rangle$ and $|\psi\rangle$ are ELO-equivalent. 
\end{theorem}
The action of an ELO on a state $|\psi\rangle $ induces Elementary Operations on submatrices of the coefficient matrix $C_{|\psi\rangle}$ of this state. Let us warm-up with the two qubit case, where an arbitrary state has coefficient matrix:
\begin{eqnarray}
C_{|\psi\rangle} = \left(\begin{array}{cc}
						{c_{00}}&{c_{01}} \\
						{c_{10}}&{c_{11}}
\end{array}\right).
\end{eqnarray} 
A single ELO on the first qubit of $|\psi\rangle$ correspond to elementary row-operations on $C_{|\psi\rangle}$. For example, $F_1(0,1)|\psi\rangle$ brings $C_{|\psi\rangle}$ to
\begin{eqnarray}
C_{F_1(0,1)|\psi\rangle}=F(0,1)C_{|\psi\rangle}=\left(\begin{array}{cc}
						{c_{01}}&{c_{11}} \\
						{c_{00}}&{c_{10}}
\end{array}\right).
\end{eqnarray}
ELO on the last qubit of $|\psi\rangle$ correspond to elementary column-operations on $C_{|\psi\rangle}$; for example, $S_2(1,\lambda)|\psi\rangle$ results in
\begin{eqnarray}
C_{S_2(1,\lambda)|\psi\rangle}=C_{|\psi\rangle}S(1,\lambda)=\left(\begin{array}{cc}
						{c_{00}}&{\lambda c_{01}} \\
						{c_{10}}&{\lambda c_{11}}
\end{array}\right).
\end{eqnarray}
In the general $d_1\otimes d_2$ bipartite case, the coefficient matrix $C_{|\psi\rangle}$ of an arbitrary state is a $d_1\times d_2$ matrix and we have 
\begin{eqnarray}
C_{E_1|\psi\rangle}=EC_{|\psi\rangle}; \ \ C_{E'_2|\psi\rangle}=C_{|\psi\rangle}E'.
\end{eqnarray}
Thus, an ELO on the first subsystem translates as application of an elementary row-operation on $C_{|\psi\rangle}$, while an ELO on the second subsystem is the same as an elementary column-operation on $C_{|\psi\rangle}$. In other words, ELO-equivalent bipartite states have coefficient matrices that are matrix equivalent.  

From the GJE method, we know that an arbitrary matrix is row-equivalent to its so-called Row-Reduced Form (RRF), also known as Row-Reduced Echelon Form. Conversely, an arbitrary matrix is column-equivalent to its Column-Reduced Form (CRF). In the bipartite case, application of the GJE strategy to the first subsystem and then further on to the second one brings the coefficient matrix to its Fully Reduced Form (FRF):
\begin{eqnarray}
\left(\begin{array}{c|c}
      {I_m}&{0_{m,d_2-m}}\\
\hline{0_{d_1-m,m}}&{0_{d_1-m,d_2-m}}
\end{array}\right)
\end{eqnarray}
where $I_m$ is the $m\times m$ identity matrix and $0_{p,n}$ denotes a $p\times n$ block composed of zeros. The number $m$ is obviously the matrix rank of $C_{|\psi\rangle}$ and is known in the literature as the Schmidt Number or Schmidt Rank of the state. As will be useful later on, one notices that $m$ is also the number of pivots of the FRF of $C_{|\psi\rangle}$. Two bipartite states are then ELO-equivalent  - and hence SLOCC-equivalent - iff they have the same Schmidt Number \cite{gtreview}, i.e., iff the FRFs of their coefficient matrices have the same number of pivots.   

\subsection{Tripartite systems}
Let us now see what an extra subsystem means for the ELO-equivalence of states. It is convenient to write the coefficient matrix in terms of square submatrices. In the three-qubit case, the coefficient matrix of a state is
\begin{eqnarray}
\left(\begin{array}{c|c}
						{A}&{B} 
					\end{array}\right) \label{3qubitCM}
\end{eqnarray} 
where $A$ and $B$ are $2\times 2$ matrices. In the three-qutrit case we have
\begin{eqnarray}
\left(\begin{array}{c|c|c}
						{A}&{B}&{C} 
					\end{array}\right)
\end{eqnarray} 
with $A$, $B$ and $C$ being $3\times 3$ matrices. The tripartite $d$-level case is then
\begin{eqnarray}
\left(\begin{array}{c|c|c|c}
						{A_0}&{A_1}&{\ldots}&{A_{d-1}} \label{3partyCM}
					\end{array}\right)
\end{eqnarray} 
where each $A_k$ is a $d\times d$ matrix. It is easy to see that
\begin{eqnarray}
C_{E_1|\psi\rangle} = \left(\begin{array}{c|c|c|c}
						{EA_0}&{EA_1}&{\ldots}&{EA_{d-1}} 
					\end{array}\right); \\
C_{E'_3|\psi\rangle} = \left(\begin{array}{c|c|c|c}
						{A_0E'}&{A_1E'}&{\ldots}&{A_{d-1}E'} 
					\end{array}\right),				
\end{eqnarray} 
i.e., an ELO acting on the first (last) qudit corresponds to an elementary collective row-operation (column-operation) on the submatrices $A_k$ of the coefficient matrix $C_{|\psi\rangle}$. Note that GJE on the first subsystem brings the whole $C_{|\psi\rangle}$ to its RRF. This is of course due to indexing the rows of the coefficient matrix by the levels of the first subsystem. GJE on the third subsystem further reduces the matrix.

What about the ``middle" subsystem? In the simplest case of three qubits (\ref{3qubitCM}), a local permutation results in
\begin{eqnarray}
\left(\begin{array}{c|c}
						{A}&{B} 
					\end{array}\right)\stackrel{F_2(0,1)}{\sim}
\left(\begin{array}{c|c}
						{B}&{A} 
					\end{array}\right).							
\end{eqnarray} 
Multiplication of level $1$ by a non-zero scalar $\lambda$ gives
\begin{eqnarray}
\left(\begin{array}{c|c}
						{A}&{B} 
					\end{array}\right)\stackrel{S_2(1,\lambda)}{\sim}
\left(\begin{array}{c|c}
						{A}&{\lambda B} 
					\end{array}\right).							
\end{eqnarray}  
In the general case (\ref{3partyCM}), an ELO on the second subsystem induces an elementary operation among the submatrices $A_1,\ldots A_{d-1}$. 
A permutation of levels gives
\begin{eqnarray}
\left(\begin{array}{c|c|c|c|c|c|c}
						{A_0}&{\ldots}&{A_i}&{\ldots}&{A_j}&{\ldots}&{A_{d-1}} 
					\end{array}\right) \\ 
					\stackrel{F_2(i,j)}{\sim}
\left(\begin{array}{c|c|c|c|c|c|c}
						{A_0}&{\ldots}&{A_j}&{\ldots}&{A_i}&{\ldots}&{A_{d-1}} 
					\end{array}\right),								
\end{eqnarray} 
while sum of a row times $\lambda$ to another row results in
\begin{eqnarray}
\left(\begin{array}{c|c|c|c|c|c|c}
						{A_0}&{\ldots}&{A_i}&{\ldots}&{A_j}&{\ldots}&{A_{d-1}} 
					\end{array}\right) \\ 
					\stackrel{L_2(i,\lambda,j)}{\sim}
\left(\begin{array}{c|c|c|c|c|c|c}
						{A_0}&{\ldots}&{A_i}&{\ldots}&{A'_j}&{\ldots}&{A_{d-1}}
					\end{array}\right);\\
						A'_j=A_j+\lambda A_i								
\end{eqnarray}     
The idea then is that applying the GJE to the three subsystems will then bring the matrix to its Multipartite Fully Reduced Form (MFRF), explained in the next section. 

Similarity transformations on the blocks can also be performed and significantly simplify the search for the MFRF of a coefficient matrix. Given that an invertible operation $S$ is a sequence of elementary matrices, $S=E^{(l)}$, we can perform a collective similarity transformation on the submatrices $A_k$ by applying $S=E^{(l)}$ to the first subsystem and $S^{-1}=(E^{(l)})^{(-1)}$ to the last subsystem:
\begin{eqnarray*}
C_{S_1S_3^{-1}|\psi\rangle} = \left(\begin{array}{c|c|c|c}
						{SA_0S^{-1}}&{SA_1S^{-1}}&{\ldots}&{SA_{d-1}S^{-1}} 
					\end{array}\right).				
\end{eqnarray*} 
As shown in the examples, in some situations the blocks $A_k$ assume the form of either the full identity operator or its restriction to a smaller subspace. Since similarity transformations do not affect the identity, one can map submatrices in other blocks into some canonical matrix decomposition, as for example Jordan normal forms. 

\subsection{N-partite states} 
The $N$-partite $d$-level coefficient matrix of an arbitrary state is given by
\begin{eqnarray}
\left(\begin{array}{c|c|c|c}
						{A_0}&{A_1}&{\ldots}&{A_n} 
					\end{array}\right)
\end{eqnarray} 
where $n=d^{N-2}-1$ and $A_k$ are $d\times d$ blocks. An ELO acting on the first and last subsystems result in
\begin{eqnarray}
C_{E_1|\psi\rangle} = \left(\begin{array}{c|c|c|c}
						{EA_0}&{EA_1}&{\ldots}&{EA_n} 
					\end{array}\right); \\
C_{E'_N|\psi\rangle} = \left(\begin{array}{c|c|c|c}
						{A_0E'}&{A_1E'}&{\ldots}&{A_nE'} 
					\end{array}\right).				
\end{eqnarray} 
ELOs on the other qudits induce linear combinations of sets of nested columns and for large $N$ there are many different possibilities. For example, the coefficient matrix of a four-qubit state $|\psi\rangle$ is given by
\begin{eqnarray}
C_{|\psi\rangle}=\left(\begin{array}{cc|cc}
      {c_{0000}}&{c_{0001}}&{c_{0010}}&{c_{0011}}\\
      {c_{1000}}&{c_{1001}}&{c_{1010}}&{c_{1011}}\\
\hline{c_{0100}}&{c_{0101}}&{c_{0110}}&{c_{0111}}\\
      {c_{1100}}&{c_{1101}}&{c_{1110}}&{c_{1111}}\end{array}\right)
\end{eqnarray}
where we have chosen a square representation for convenience. In terms of $2\times 2$ blocks, we have
\begin{eqnarray}
C_{|\psi\rangle}=\left(\begin{array}{c|c}
      {A}&{B}\\
\hline{C}&{D}\end{array}\right)
\end{eqnarray}
The action of an ELO on the subsystems result in
\begin{eqnarray}
C_{E_1|\psi\rangle} &=&\left(\begin{array}{c|c}
      {EA}&{EB}\\
\hline{EC}&{ED}\end{array}\right); \\
C_{L_2(0,\lambda,1)|\psi\rangle} &=&\left(\begin{array}{c|c}
      {A}&{B}\\
\hline{C+\lambda A}&{D+\lambda B}\end{array}\right); \\
C_{L_3(0,\lambda,1)|\psi\rangle} &=&\left(\begin{array}{c|c}
      {A}&{B+\lambda A}\\
\hline{C}&{D+ \lambda C}\end{array}\right); \\
C_{E_4|\psi\rangle} &=&\left(\begin{array}{c|c}
      {AE}&{BE}\\
\hline{CE}&{DE}\end{array}\right).
\end{eqnarray}\newline
In the standard GJE procedure one reduces a matrix to its RRF via Elementary Row-Operations. The RRF of a matrix is unique and is in great part characterized by the so-called pivots, which have unique relative positions to one another. The other aspect that characterizes the RRF is the coefficients other than pivots, the free parameters. If there are no free parameters, the corresponding system of linear equations has unique solution. For example, if the coefficient matrix of the system is invertible, its RRF is the identity matrix and the solution is unique. Whenever free parameters are present, the corresponding system of linear equations - if consistent - has infinite solutions, since each different value the free parameter can assume will correspond to a different solution. Nevertheless, the coefficient matrix of the system has unique RRF, i.e., the coefficients appearing in the RRF are unique. 

Considering now coefficient matrices of multipartite states, it was shown previously that besides Elementary Row-Operations one can perform other types of collective Elementary Operations. In the bipartite case, the whole set of Elementary Column-Operations is available as well. In this case, the GJE procedure brings the coefficient matrix to its FRF, which is unique (up to local permutations of levels). Moreover, there are no free-parameters, only pivots. Hence, the number of pivots uniquely characterizes the SLOCC-equivalence of a bipartite state. 

The general multipartite case is more complicated to characterize in this respect. The first difference is the concept of pivots. In the bipartite case, the RRF of the coefficient matrix already determine the pivots. In the multipartite case, as shown in some examples of this article, the RRF is just the starting point. What first appears as a pivot in the RRF might be removed later on by ELOs during the GJE. The second difference is the presence of free parameters, which imply infinite SLOCC-inequivalent classes - and how to identify them.

The overall idea to characterize the SLOCC-equivalence of multipartite states is to bring a coefficient matrix to its MFRF, the unique Reduced Form achieved via the action of ELOs that correspond to the GJE method. Two states are then SLOCC-equivalent iff their MFRFs are identical. The SLOCC classification problem then becomes the problem of classification of different MFRFs, which would constitute a considerable simplification. 

The full classification of MFRFs is a challenging problem and will be worked elsewhere. The examples worked in what follows mostly show SLOCC-equivalences to well-known classes of multipartite states, such as GHZ-states, while inequivalent classes are simple to discriminate. 

\textit{Note:} The normalization of a state is irrelevant for its SLOCC-equivalence. Hence in what follows we leave states un-normalized whenever convenient.  

\subsection{Example: Three-qubit case}
Let us re-work the well-known three-qubit case, now from the point of view of ELOs. As stated previously, the coefficient matrix of an arbitrary three-qubit state $|\psi\rangle$ reads: 
\begin{eqnarray}
C_{|\psi\rangle} = \left(\begin{array}{c|c}
						{A}&{B} 
					\end{array}\right)
\end{eqnarray} 
where $A$ and $B$ are $2\times 2$ blocks. GJE on the first qubit will bring $C_{|\psi\rangle}$ to its RRF. The first possibility is:
\begin{eqnarray*}
\left(\begin{array}{cc|cc}
						{1}&{a}&{b}&{c} \\
						{0}&{0}&{0}&{0}
\end{array}\right).
\end{eqnarray*} 
Using the pivot, we eliminate the scalar $a$ by applying $L_3(0,-a,1)|\psi\rangle$:
\begin{eqnarray*}
\left(\begin{array}{cc|cc}
						{1}&{0}&{b}&{c'} \\
						{0}&{0}&{0}&{0}
\end{array}\right).
\end{eqnarray*} 
where $c'=c-a$. Applying $L_2(0,-b,1)$ gives:
\begin{eqnarray*}
\left(\begin{array}{cc|cc}
						{1}&{0}&{0}&{c'} \\
						{0}&{0}&{0}&{0}
\end{array}\right).
\end{eqnarray*}
If $c=a$, $c'=0$ and the MFRF corresponds to $|000\rangle$, a full separable state. If $c\neq a$, applying $S_3(1,1/c')$ results in the MFRF:
\begin{eqnarray*}
\left(\begin{array}{cc|cc}
						{1}&{0}&{0}&{1} \\
						{0}&{0}&{0}&{0}
\end{array}\right),
\end{eqnarray*} 
which corresponds to $|0\rangle(|00\rangle+|11\rangle)$, a biseparable state.\newline
The second possible RRF is 
\begin{eqnarray*}
\left(\begin{array}{cc|cc}
						{1}&{a}&{b}&{0} \\
						{0}&{0}&{0}&{1}
\end{array}\right).
\end{eqnarray*} 
Applying the same sequence of ELOs, $S_3(1,1/c')L_2(0,-b,1)L_3(0,-a,1)|\psi\rangle$ gives us the MFRF
\begin{eqnarray*}
\left(\begin{array}{cc|cc}
						{1}&{0}&{0}&{0} \\
						{0}&{0}&{0}&{1}
\end{array}\right).
\end{eqnarray*}
which is $|000\rangle+|111\rangle$, the GHZ-state.\newline
The third possible RRF is
\begin{eqnarray*}
\left(\begin{array}{cc|cc}
						{1}&{a}&{0}&{b} \\
						{0}&{0}&{1}&{c}
\end{array}\right).
\end{eqnarray*}
The submatrix $\left(\begin{array}{cc}{1}&{0}\\{0}&{1}\end{array}\right)$ is the identity, while $\left(\begin{array}{cc}{a}&{b}\\{0}&{c}\end{array}\right)$ is diagonalized by $N=\left(\begin{array}{cc}{1}&{b/\eta}\\{0}&{(c-a)/\eta}\end{array}\right)$, $\eta=[b^2+(c-a)^2]^{1/2}$. Thus applying $N_1N_2^{-1}$ results in
\begin{eqnarray*}
\left(\begin{array}{cc|cc}
						{1}&{a}&{0}&{0} \\
						{0}&{0}&{1}&{c}
\end{array}\right).
\end{eqnarray*}
Applying $L_3(0,-a,1)$ in order to eliminate $a$ gives
\begin{eqnarray*}
\left(\begin{array}{cc|cc}
						{1}&{0}&{0}&{0} \\
						{0}&{0}&{1}&{0}
\end{array}\right),
\end{eqnarray*}
if $c=a$ and the MFRF is $(|00\rangle+|11\rangle)|0\rangle$, biseparable in this case; if $c\neq a$, we apply $L_3(1,-1,0)S_3(1,(c-a)^{-1})$, obtaining  
\begin{eqnarray*}
\left(\begin{array}{cc|cc}
						{1}&{0}&{0}&{0} \\
						{0}&{0}&{0}&{1}
\end{array}\right)
\end{eqnarray*}
the MFRF $|000\rangle+|111\rangle$, the GHZ-state.

The last possible RRF is
\begin{eqnarray*}
\left(\begin{array}{cc|cc}
						{1}&{0}&{a}&{b} \\
						{0}&{1}&{c}&{d}
\end{array}\right).
\end{eqnarray*} 
The first block is a $2\times 2$ identity block. We can thus apply the similarity transformation $S$ that brings the block 
$M=\left(\begin{array}{cc}
						{a}&{b} \\
						{c}&{d}
\end{array}\right)$ to its Jordan normal form $SMS^{-1}$. Thus we apply $S_1S_3|\psi\rangle$, obtaining either $\left(\begin{array}{cc}
						{\lambda}&{1} \\
						{0}&{\lambda}
\end{array}\right)$ or $\left(\begin{array}{cc}
						{\lambda_1}&{0} \\
						{0}&{\lambda_2}
\end{array}\right)$. The first possibility results in:
\begin{eqnarray*}
\left(\begin{array}{cc|cc}
						{1}&{0}&{\lambda}&{1} \\
						{0}&{1}&{0}&{\lambda}
\end{array}\right)\\
\stackrel{L_2(0,-\lambda,1)}{\sim}\left(\begin{array}{cc|cc}
						{1}&{0}&{0}&{1} \\
						{0}&{1}&{0}&{0}
\end{array}\right)\\
\stackrel{F_3(0,1)}{\sim}\left(\begin{array}{cc|cc}
						{0}&{1}&{1}&{0} \\
						{1}&{0}&{0}&{0}
\end{array}\right) 
\end{eqnarray*}
with the MFRF of $|100\rangle+|010\rangle+|001\rangle$, the W-state; the second possibility gives
\begin{eqnarray*}
\left(\begin{array}{cc|cc}
						{1}&{0}&{\lambda_1}&{0} \\
						{0}&{0}&{0}&{\lambda_2}
\end{array}\right)\\
\stackrel{L_2(0,-\lambda_1,1)}{\sim}\left(\begin{array}{cc|cc}
						{1}&{0}&{0}&{0} \\
						{0}&{0}&{0}&{\lambda_2}
\end{array}\right)\\
\stackrel{S_3(1,\lambda_2^{-1})}{\sim}\left(\begin{array}{cc|cc}
						{1}&{0}&{0}&{0} \\
						{0}&{0}&{0}&{1}
\end{array}\right),
\end{eqnarray*} 
with MFRF of $|000\rangle+|111\rangle$, the GHZ-state. The special cases of scalars with identical or null values were excluded, for simplicity and are easily worked out. 

The different MFRFs found are not SLOCC-equivalent through a simple observation: ELOs do not increase the ranks of submatrices of the coefficient matrix (e.g. \cite{dafa}). For example, the GHZ-state and the W-state are not SLOCC-equivalent because the ranks of their respective submatrices $\left(\begin{array}{cc}{0}&{0}\\{0}&{1}\end{array}\right)$ and $\left(\begin{array}{cc}{0}&{1}\\{1}&{0}\end{array}\right)$ are different, even though the Row-Rank of the their coefficient matrix is the same. 

\subsection{Example: Three qutrits, special case}

The coefficient matrix of an arbitrary three-qutrit state is given by
\begin{eqnarray*}
\left(
\begin{array}{ccc|ccc|ccc}
						{c_{000}}&{c_{001}}&{c_{002}}&{c_{010}}&{c_{011}}&{c_{012}}&{c_{020}}&{c_{021}}&{c_{022}}\\
						{c_{100}}&{c_{101}}&{c_{102}}&{c_{110}}&{c_{111}}&{c_{112}}&{c_{120}}&{c_{121}}&{c_{122}}\\
						{c_{200}}&{c_{201}}&{c_{202}}&{c_{210}}&{c_{211}}&{c_{212}}&{c_{220}}&{c_{221}}&{c_{222}}
						\end{array}\right).
\end{eqnarray*}
Let us consider here a simple special case, where GJE on the first qutrit gives the following RRF:
\begin{eqnarray*}
\left(
\begin{array}{ccc|ccc|ccc}
						{1}&{a}&{b}&{c}&{0}&{d}&{e}&{f}&{0}\\
						{0}&{0}&{0}&{0}&{1}&{g}&{h}&{i}&{0}\\
						{0}&{0}&{0}&{0}&{0}&{0}&{0}&{0}&{1}
						\end{array}\right)
\end{eqnarray*}						
Using the first pivot, we eliminate scalars $a$ and $b$ via $L_3(0,-b,2)L_3(0,-a,1)|\psi\rangle$: 						
\begin{eqnarray*}
\left(
\begin{array}{ccc|ccc|ccc}
						{1}&{0}&{0}&{c}&{c'}&{c''}&{e}&{e'}&{e''}\\
						{0}&{0}&{0}&{0}&{1}&{g}&{h}&{h'}&{h''}\\
						{0}&{0}&{0}&{0}&{0}&{0}&{0}&{0}&{1}
						\end{array}\right)
\end{eqnarray*}
with obvious notation. Using the second pivot we eliminate $g$ via $L_3(1,-g,2)$:
\begin{eqnarray*}
\left(
\begin{array}{ccc|ccc|ccc}
						{1}&{0}&{0}&{c}&{c'}&{c'''}&{e}&{e'}&{e'''}\\
						{0}&{0}&{0}&{0}&{1}&{0}&{h}&{h'}&{h'''}\\
						{0}&{0}&{0}&{0}&{0}&{0}&{0}&{0}&{1}
						\end{array}\right)
\end{eqnarray*}
Using the second and third pivots, we eliminate scalars $c'$, $h'''$ and $e'''$ via $L_1(1,-c',0)L_1(2,-h''',1)L_1(2,-e''',0)$: 
\begin{eqnarray*}
\left(
\begin{array}{ccc|ccc|ccc}
						{1}&{0}&{0}&{c}&{0}&{c'''}&{j}&{j'}&{0}\\
						{0}&{0}&{0}&{0}&{1}&{0}&{h}&{h'}&{0}\\
						{0}&{0}&{0}&{0}&{0}&{0}&{0}&{0}&{1}
						\end{array}\right)
\end{eqnarray*}
Using the first and second pivots, we eliminate escalars $c$, $j$ and $h'$ via $L_2(1,-h',2)L_2(0,-j,2)L_2(0,-c,1)|\psi\rangle$: 
\begin{eqnarray*}
\left(
\begin{array}{ccc|ccc|ccc}
						{1}&{0}&{0}&{0}&{0}&{c'''}&{0}&{j'}&{-c'''}\\
						{0}&{0}&{0}&{0}&{1}&{0}&{h}&{0}&{0}\\
						{0}&{0}&{0}&{0}&{0}&{0}&{0}&{0}&{1}
						\end{array}\right)
\end{eqnarray*}
Using the third pivot, we eliminate escalar $-c'''$ via $L_1(2,c''',0)$. We now normalize the scalars $h$, $j'$ and $c'''$ via $S_2(0,hj')S_3(0,(hj')^{-1})S_1(1,j')S_3(1,(j')^{-1})\\ S_1(2,c''')S_3(2,(c''')^{-1})$:
\begin{eqnarray*}
\left(
\begin{array}{ccc|ccc|ccc}
						{1}&{0}&{0}&{0}&{0}&{1}&{0}&{1}&{0}\\
						{0}&{0}&{0}&{0}&{1}&{0}&{1}&{0}&{0}\\
						{0}&{0}&{0}&{0}&{0}&{0}&{0}&{0}&{1}
						\end{array}\right)
\end{eqnarray*}
and the final MFRF corresponds to the state $|000\rangle+|111\rangle+|222\rangle+|012\rangle+|021\rangle$.

\section{Applications: LME and hypergraph states}

Let us now apply our methods to the classes of LME and hypergraph states, which have simple coefficient matrices. For our purposes, it is enough to describe each example by their coefficient matrices; the interested reader can consult \cite{hypergraphpapers,hyper1,hyper2}.

\subsection{Elementary LME and hypergraph states}

An elementary N-qubit LME state is one with the following coefficient matrix:
\begin{eqnarray}
C_{|\phi\rangle}=\left(
\begin{array}{cc|cc|c|cc}
						{1}&{1}&{1}&{1}&{\ldots}&{1}&{1}\\
						{1}&{1}&{1}&{1}&{\ldots}&{1}&{e^{i\phi}}
\end{array}
\right)
\end{eqnarray}
i.e., all coefficients equal to $1$ except the one multiplying $|11\ldots 1\rangle$, which has an arbitrary phase $e^{i\phi}$. When $\phi=\pi$, the state is called an elementary hypergraph state. 
\begin{observation}
An elementary $N$-qubit LME state $|\phi\rangle$ is SLOCC-equivalent to the $N$-partite GHZ-state ($|0\rangle^{\otimes N}+|1\rangle^{\otimes N}$).
\end{observation}
\textbf{Proof:}
Instead of using coefficient matrices, it is more convenient to write the state explicitly and use the GJE strategy in each step. An elementary LME state $|\phi\rangle$ in terms of the the computational basis is given by: 
\begin{eqnarray*}
|\phi\rangle &=& |0\rangle\otimes\sum_{i_2\ldots i_N}|{i_2\ldots i_N}\rangle \\
	  &+& |1\rangle\otimes\sum_{i_2\ldots i_M}|{i_2\ldots i_N}\rangle\\
    &+&(e^{i\phi}-1)|111\ldots 1\rangle
\end{eqnarray*}
Applying $S_1(1,(e^{i\phi}-1)^{-1})L_1(0,-1,1)$ we obtain 
\begin{eqnarray*}
 &\sim& |00\rangle\otimes\sum_{i_3\ldots i_N}|{i_3\ldots i_N}\rangle \\
	  &+& |01\rangle\otimes\sum_{i_3\ldots i_N}|{i_3\ldots i_N}\rangle\\
		&+& |111\ldots 1\rangle
\end{eqnarray*}
Applying $L_2(0,-1,1)$ we obtain
\begin{eqnarray*}
 &\sim& |000\rangle\otimes\sum_{i_4\ldots i_N}|{i_4\ldots i_N}\rangle \\
	  &+& |001\rangle\otimes\sum_{i_4\ldots i_N}|{i_4\ldots i_N}\rangle\\
		&+& |111\ldots 1\rangle
\end{eqnarray*}
By induction, applying $L_N(0,-1,1),\ldots, L_4(0,-1,1)L_3(0,-1,1)$ will bring the state into the GHZ-state $|0\rangle^{\otimes N}+|1\rangle^{\otimes N}$. q.e.d..

\subsection{$4$-partite qubit hypergraph states}

In \cite{hyper1}, the full LU-classification of $4$-qubit hypergraph states was performed and $27$ LU-inequivalent classes were found. We show that states $|V_1\rangle$, $|V_4\rangle$ and $|V_{18}\rangle$ in that article are SLOCC-equivalent to the state with coefficient matrix
\begin{eqnarray}
C_{|\mu\rangle}=\left(\begin{array}{cc|cc}
      {1}&{0}&{0}&{0}\\
      {0}&{0}&{0}&{0}\\
\hline{0}&{0}&{0}&{0}\\
      {1}&{0}&{0}&{1}
\end{array}
\right)
\end{eqnarray}
i.e., $|\mu\rangle=|0000\rangle+|1100\rangle+|1111\rangle$.

State $|V_{18}\rangle$ has coefficient matrix 
\begin{eqnarray}
C_{|V_{18}\rangle}=\left(\begin{array}{cc|cc}
      {1}&{1}&{1}&{1}\\
      {1}&{1}&{1}&{1}\\
\hline{1}&{1}&{1}&{1}\\
      {-1}&{-1}&{-1}&{1}
\end{array}
\right)
\end{eqnarray}
Performing the sequence of ELOs $S_4(1,-1)L_2(0,-1,1)L_1(1,1,0)L_3(0,-1,1)L_4(0,-1,1)\\ L_1(1,-1,0)S_1(1,-1/2)L_1(0,-1,1)|V_{18}\rangle$, results in $C_{|V_{18}\rangle}$ being mapped into $C_{|\mu\rangle}$. 

State $|V_1\rangle=|0000\rangle+|0001\rangle+|1100\rangle+|1111\rangle$ has coefficient matrix
\begin{eqnarray}
C_{|V_1\rangle}=\left(\begin{array}{cc|cc}
      {1}&{1}&{0}&{0}\\
      {0}&{0}&{0}&{0}\\
\hline{0}&{0}&{0}&{0}\\
      {1}&{0}&{0}&{1}
\end{array}
\right)
\end{eqnarray}
Performing the sequence of ELOs $L_3(1,1,0)L_4(0,-1,1)|V_1\rangle$, results in $C_{|V_1\rangle}$ being mapped into $C_{|\mu\rangle}$. 

State $|V_4\rangle=|0000\rangle+|0001\rangle+|0010\rangle+|1111\rangle$ has coefficient matrix
\begin{eqnarray}
C_{|V_4\rangle}=\left(\begin{array}{cc|cc}
      {1}&{1}&{1}&{0}\\
      {0}&{0}&{0}&{0}\\
\hline{0}&{0}&{0}&{0}\\
      {0}&{0}&{0}&{1}
\end{array}
\right)
\end{eqnarray}
Performing the sequence of ELOs $F_1(0,1)F_4(0,1)F_3(0,1)F_2(0,1)S_1(1,-1)S_4(1,-1)\\ L_3(0,-1,1)L_4(0,-1,1)|V_4\rangle$, results in $C_{|V_4\rangle}$ being mapped into $C_{|\mu\rangle}$. 

\subsection{Tripartite qudit hypergraphs}

In \cite{hyper2}, the SLOCC and LU classification of tripartite qutrits and ququarts hypergraph states was achieved. There it was found by numerical search that the elementary three-qutrit hypergraph state is SLOCC-equivalent to the three-qutrit GHZ-state, while there was strong numerical evidence - but no analytical demonstration -that the elementary three-ququart hypergraph state is not SLOCC-equivalent to the three-ququart GHZ-state.  

The coefficient matrix of a three-qutrit elementary hypergraph $|H_3\rangle$is given by  
\begin{eqnarray}
\left(
\begin{array}{ccc|ccc|ccc}
						{1}&{1}&{1}&{1}&{1}&{1}&{1}&{1}&{1}\\
						{1}&{1}&{1}&{1}&{\omega}&{\omega^2}&{1}&{\omega^2}&{\omega}\\
						{1}&{1}&{1}&{1}&{\omega^2}&{\omega}&{1}&{\omega}&{\omega^2}
\end{array}
\right)
\end{eqnarray}
where $\omega=e^{i(2\pi/3)}$ is the 3rd root of unity $1+\omega+\omega^2=0$. The vectors $\{|p_0\rangle=(1,1,1),|p_1\rangle=(1,\omega,\omega^2),|p_2\rangle=(1,\omega^2,\omega)\}$ form an orthogonal basis of $\mathbb{C}^3$, satisfying $|p_i\rangle=F|i\rangle$, where $F=\sqrt{1/d}\sum_{m,n} \omega^{mn}|m\rangle\langle n|$ is the discrete Fourier transform \cite{vourdas}. Bringing these vectors back to the canonical basis $\{(1,0,0),(0,1,0),(0,0,1)\}$ by $F_1^{-1}|H_3\rangle$, we get:
\begin{eqnarray}
\left(
\begin{array}{ccc|ccc|ccc}
						{1}&{1}&{1}&{1}&{0}&{0}&{1}&{0}&{0}\\
						{0}&{0}&{0}&{0}&{1}&{0}&{0}&{0}&{1}\\
						{0}&{0}&{0}&{0}&{0}&{1}&{0}&{1}&{0}
\end{array}
\right).
\end{eqnarray}
Further reduction through $L_1(2,1,0)L_1(1,1,0)L_1(2,1,0)L_2(0,-1,2)L_2(0,-1,1)$ yields:
\begin{eqnarray}
\left(
\begin{array}{ccc|ccc|ccc}
						{1}&{0}&{0}&{0}&{0}&{0}&{0}&{0}&{0}\\
		      	{0}&{0}&{0}&{0}&{1}&{0}&{0}&{0}&{1}\\
						{0}&{0}&{0}&{0}&{0}&{1}&{0}&{1}&{0}
\end{array}
\right).
\end{eqnarray}
The second block has an identity on the subspace spanned by $\{|1\rangle,|2\rangle\}$, while the third block is the Pauli $\sigma_x$ on the same subspace; we can thus apply $U_1U^{-1}_3$, where 
\begin{eqnarray}
U=2^{-1/2}\left(\begin{array}{c|cc}
{2^{1/2}}&{0}&{0}\\
\hline{0}&{1}&{1}\\
{0}&{1}&{-1}
\end{array}\right),
\end{eqnarray}
i.e., the Hadamard on the subspace spanned by $\{|1\rangle,|2\rangle\}$, and the coefficient matrix is mapped into 
\begin{eqnarray}
\left(
\begin{array}{ccc|ccc|ccc}
						{1}&{0}&{0}&{0}&{0}&{0}&{0}&{0}&{0}\\
		      	{0}&{0}&{0}&{0}&{1}&{0}&{0}&{1}&{0}\\
						{0}&{0}&{0}&{0}&{0}&{1}&{0}&{0}&{-1}
\end{array}
\right)
\end{eqnarray}
Applying $S_1(2,-1/2)L_2(2,1/2,1)L_2(1,-1,2)$ we get the MFRF:
\begin{eqnarray}
\left(
\begin{array}{ccc|ccc|ccc}
						{1}&{0}&{0}&{0}&{0}&{0}&{0}&{0}&{0}\\
		      	{0}&{0}&{0}&{0}&{1}&{0}&{0}&{0}&{0}\\
						{0}&{0}&{0}&{0}&{0}&{0}&{0}&{0}&{1}
\end{array}
\right)
\end{eqnarray} 
the three-qutrit GHZ-state $|000\rangle+|111\rangle+|222\rangle$.

The three-ququart elementary hypergraph state $|H_4\rangle$ has coefficient matrix   
\begin{eqnarray*}
\left(
\begin{array}{cccc|cccc|cccc|cccc}
						{1}&{1}&{1}&{1}&{1}&{1}&{1}&{1}&{1}&{1}&{1}&{1}&{1}&{1}&{1}&{1}\\
		      	{1}&{1}&{1}&{1}&{1}&{\omega}&{\omega^2}&{\omega^3}&{1}&{\omega^2}&{1}&{\omega^2}&{1}&{\omega^3}&{\omega^2}&{\omega}\\
						{1}&{1}&{1}&{1}&{1}&{\omega^2}&{1}&{\omega^2}&{1}&{1}&{1}&{1}&{1}&{\omega^2}&{1}&{\omega^2}\\
						{1}&{1}&{1}&{1}&{1}&{\omega^3}&{\omega^2}&{\omega}&{1}&{\omega^2}&{1}&{\omega^2}&{1}&{\omega}&{\omega^2}&{\omega^3}
\end{array}
\right)
\end{eqnarray*} 
where now $\omega=i$ is the fourth root of unity: $1+\omega+\omega^2+\omega^3=0$. The vectors $\{|p_0\rangle=(1,1,1,1),|p_1\rangle=(1,\omega,\omega^2,\omega^3),|p_2\rangle=(1,\omega^2,1,\omega^2),|p_3\rangle=(1,\omega^3,\omega^2,\omega)\}$ form an orthogonal basis of $\mathbb{C}^4$. Adopting the same procedure of the previous case, we apply $F_1^{-1}|H_4\rangle$, which changes this basis into the canonical basis, obtaining
\begin{eqnarray}
\left(
\begin{array}{cccc|cccc|cccc|cccc}
						{1}&{1}&{1}&{1}&{1}&{0}&{0}&{0}&{1}&{0}&{1}&{0}&{1}&{0}&{0}&{0}\\
		      	{0}&{0}&{0}&{0}&{0}&{1}&{0}&{0}&{0}&{0}&{0}&{0}&{0}&{0}&{0}&{1}\\
						{0}&{0}&{0}&{0}&{0}&{0}&{1}&{0}&{0}&{1}&{0}&{1}&{0}&{0}&{1}&{0}\\
						{0}&{0}&{0}&{0}&{0}&{0}&{0}&{1}&{0}&{0}&{0}&{0}&{0}&{1}&{0}&{0}
\end{array}
\right).
\end{eqnarray} 
Applying $L_1(3,1,0)L_1(2,1,0)L_1(1,1,0)L_2(0,-1,3)\\ L_2(0,-1,2)L_2(0,-1,1)$ gives us
\begin{eqnarray}
\left(
\begin{array}{cccc|cccc|cccc|cccc}
						{1}&{0}&{0}&{0}&{0}&{0}&{0}&{0}&{0}&{0}&{0}&{0}&{0}&{0}&{0}&{0}\\
		      	{0}&{0}&{0}&{0}&{0}&{1}&{0}&{0}&{0}&{0}&{0}&{0}&{0}&{0}&{0}&{1}\\
						{0}&{0}&{0}&{0}&{0}&{0}&{1}&{0}&{0}&{1}&{0}&{1}&{0}&{0}&{1}&{0}\\
						{0}&{0}&{0}&{0}&{0}&{0}&{0}&{1}&{0}&{0}&{0}&{0}&{0}&{1}&{0}&{0}
\end{array}
\right)
\end{eqnarray}
By the same rationale of the previous case, we apply $U_1U^{-1}_3$, where
\begin{eqnarray}
U=2^{-1/2}\left(\begin{array}{c|ccc}
{2^{1/2}}&{0}&{0}&{0}\\
\hline{0}&{1}&{0}&{1}\\
{0}&{0}&{\sqrt{2}}&{0}\\
{0}&{1}&{0}&{-1}
\end{array}\right)
\end{eqnarray}
obtaining 
\begin{eqnarray}
\left(
\begin{array}{cccc|cccc|cccc|cccc}
						{1}&{0}&{0}&{0}&{0}&{0}&{0}&{0}&{0}&{0}&{0}&{0}&{0}&{0}&{0}&{0}\\
		      	{0}&{0}&{0}&{0}&{0}&{1}&{0}&{0}&{0}&{0}&{0}&{0}&{0}&{1}&{0}&{0}\\
						{0}&{0}&{0}&{0}&{0}&{0}&{1}&{0}&{0}&{\sqrt{2}}&{0}&{0}&{0}&{0}&{1}&{0}\\
						{0}&{0}&{0}&{0}&{0}&{0}&{0}&{1}&{0}&{0}&{0}&{0}&{0}&{0}&{0}&{-1}
\end{array}
\right).
\end{eqnarray}
Applying $S_2(3,-1/2)L_2(3,-1/2,1)L_2(1,-1,3)S_2(2,2^{-1/2})$ and the MFRF is 
\begin{eqnarray}
\left(
\begin{array}{cccc|cccc|cccc|cccc}
						{1}&{0}&{0}&{0}&{0}&{0}&{0}&{0}&{0}&{0}&{0}&{0}&{0}&{0}&{0}&{0}\\
		      	{0}&{0}&{0}&{0}&{0}&{1}&{0}&{0}&{0}&{0}&{0}&{0}&{0}&{0}&{0}&{0}\\
						{0}&{0}&{0}&{0}&{0}&{0}&{1}&{0}&{0}&{1}&{0}&{0}&{0}&{0}&{0}&{0}\\
						{0}&{0}&{0}&{0}&{0}&{0}&{0}&{0}&{0}&{0}&{0}&{0}&{0}&{0}&{0}&{1}
\end{array}
\right)
\end{eqnarray}
which is the state $|000\rangle+|111\rangle+|2\rangle(|12\rangle+|21\rangle)+|333\rangle$ obviously not equivalent to the GHZ-state, as the numerical evidence of \cite{hyper2} suggested.

\section{Remarks on Dicke-like structures}

Let us consider a special class of states where the coefficients in the expansion (\ref{expansion}) satisfy $c_{i_1i_2\ldots i_N}=c_{i_1+i_2+\ldots i_N}$. A state in this class will have coefficient matrix with Hankel form (see discussion in \cite{becv} for a parallel formulation in quantum optics). 
\begin{eqnarray*}
\left(
\begin{array}{cccc|cccc|cccc}
						{c_0}&{c_1}&{c_2}&{\ldots}&{c_1}&{c_2}&{c_3}&{\ldots}&{}&{}&{}&{}\\
		      	{c_1}&{c_2}&{c_3}&{\ldots}&{c_2}&{c_3}&{c_4}&{\ldots}&{}&{\ldots}&{\ldots}&{}\\
						{c_2}&{c_3}&{c_4}&{\ldots}&{c_3}&{c_4}&{c_5}&{\ldots}&{}&{}&{}&{}\\
						{\vdots}&{\vdots}&{\vdots}&{\ddots}&{\vdots}&{\vdots}&{\vdots}&{\ddots}&{}&{}
\end{array}
\right).
\end{eqnarray*} 
For example, a three-qubit state in this class have coefficient matrix given by   
\begin{eqnarray}
\left(
\begin{array}{cc|cc}
						{c_0}&{c_1}&{c_1}&{c_2}\\
		      	{c_1}&{c_2}&{c_2}&{c_3}						
\end{array}
\right)
\end{eqnarray} 
Taking $c_0=1$ and the other $c_k$'s zero, we have the ``vacuum" $|000\rangle$; for $c_1=1$ and the others zero, we have the W-state $|100\rangle+|010\rangle+|001\rangle$. But even if both are non-zero, the state would be ELO-equivalent to the W-state:  
\begin{eqnarray}
\left(
\begin{array}{cc|cc}
						{c_0}&{c_1}&{c_1}&{0}\\
		      	{c_1}&{0}&{0}&{0}						
\end{array}
\right)\stackrel{L_1(1,-c_0/c_1,0)}{\sim}\left(
\begin{array}{cc|cc}
						{0}&{c_1}&{c_1}&{0}\\
		      	{c_1}&{0}&{0}&{0}						
\end{array}
\right),
\end{eqnarray}
which is the W-state times $c_0$. However, if $c_0,c_1,c_2\neq 0$, this would be no longer valid. 

In the four-qubit case, the coefficient matrix of a state in this class is given by
\begin{eqnarray}
\left(\begin{array}{cc|cc}
      {c_0}&{c_1}&{c_1}&{c_2}\\
      {c_1}&{c_2}&{c_2}&{c_3}\\
\hline{c_1}&{c_2}&{c_2}&{c_3}\\
      {c_2}&{c_3}&{c_3}&{c_4}\end{array}\right).
\end{eqnarray}
It is easy to check that the same applies: if $c_0$ and $c_1$ are non-null while the rest are null, the state is ELO-equivalent to the W-state $|1000\rangle+|0100\rangle+|0010\rangle+|0001\rangle$, while if any other $c_k$ is non-null the ELO-equivalence to the W-state is not guaranteed. 

The coefficient matrix of a three-qutrit state in this class is given by
\begin{eqnarray*}
\left(
\begin{array}{ccc|ccc|ccc}
						{c_0}&{c_1}&{c_2}&{c_1}&{c_2}&{c_3}&{c_2}&{c_3}&{c_4}\\
		      	{c_1}&{c_2}&{c_3}&{c_2}&{c_3}&{c_4}&{c_3}&{c_4}&{c_5}\\
						{c_2}&{c_3}&{c_4}&{c_3}&{c_4}&{c_5}&{c_4}&{c_5}&{c_6}
\end{array}
\right)
\end{eqnarray*} 
Here there is a similar situation, but for the more general situation where coefficients $c_0,c_1,c_2\neq 0$, while the others are zero. In this case, we have
\begin{eqnarray}
\left(
\begin{array}{ccc|ccc|ccc}
						{c_0}&{c_1}&{1}&{c_1}&{1}&{0}&{1}&{0}&{0}\\
		      	{c_1}&{1}&{0}&{1}&{0}&{0}&{0}&{0}&{0}\\
						{1}&{0}&{0}&{0}&{0}&{0}&{0}&{0}&{0}
\end{array}
\right)
\end{eqnarray} 
where we made $c_2=1$, without loss of generality. Applying $L_1(1,-c_1,0)L_1(2,-c_0,0)L_1(2,-c_1,1)$, the coefficient matrix has MFRF
\begin{eqnarray}
\left(
\begin{array}{ccc|ccc|ccc}
						{0}&{0}&{1}&{0}&{1}&{0}&{1}&{0}&{0}\\
		      	{0}&{1}&{0}&{1}&{0}&{0}&{0}&{0}&{0}\\
						{1}&{0}&{0}&{0}&{0}&{0}&{0}&{0}&{0}
\end{array}
\right)
\end{eqnarray} 
which corresponds to $|200\rangle+|020\rangle+|110\rangle+|101\rangle+|011\rangle+|002\rangle$, a full Dicke-like state with $2$ excitations. If, on the other hand, any other $c_k$ is non-null along with $c_0,c_1,c_2$, the ELO-equivalence class of the corresponding state is not necessarily a Dicke-like state. Hence, for states possessing the Dicke-like structure here outlined, it appears that there is a maximal excitation (some sort of ``energy cutoff"') related to the local dimensions which eliminates all smaller excitations contributions, in the sense that these smaller excitations are irrelevant for the SLOCC-equivalence of the state. 

In multipartite infinite-dimensional systems, however, there is no bound on local dimensions and we conclude that superpositions of Dicke-like states are equivalent to the Dicke-like state with highest excitation. However, in a more general situation where the state is an infinite superposition of Dicke-like states, the coefficients $c_0,c_1,c_2,\ldots$ constitute an infinite sequence and the SLOCC-equivalence of the state is very likely determined by this sequence. A simple example is the geometric sequence $c_k=q^k$; it  corresponds to a coefficient matrix with rank 1, i.e., not a genuinely multipartite entangled state. As in \cite{becv}, in the general case there seems to exist a relation between SLOCC-equivalence and the so-called Problem of Moments \cite{shohat}, where a sequence of coefficients is determined by the moments of a measure if the Hankel matrix constructed from these coefficients is positive semidefinite. These interesting considerations are left for future work. 

\section{Conclusions and perspectives} 

In this work a method was developed for verifying the SLOCC-equivalence of multipartite pure states based on the action of ELOs. A multipartite analogue of the GJE procedure was constructed and various examples of SLOCC-equivalences and inequivalences were shown. Specifically, elementary multi-qubit LME-states were shown to be SLOCC-equivalent to the GHZ-state and results obtained numerically for qudit hypergraphs were here proved analytically. A relationship between the local dimension and SLOCC-equivalence of superpositions of Dicke-like states was hinted and it might have interesting consequences in future works connecting dimensionality, energy and entanglement. Moreover, ELOs might be considerably useful in detection schemes such as \cite{pptmixtures,sloccwit}, which are formulated in terms of the SLOCC-equivalence classes of pure multipartite representatives. Hence, the efficient detection of genuinely multipartite entanglement of large sets of mixed states depends on the SLOCC-classification of pure multipartite states \cite{note}.  since it is a special case of a local filtering operation \cite{multientang} and thus not a general SLOCC map.

The use of ELOs has the major advantage of giving constructive analytical demonstrations of SLOCC-equivalences or inequivalences. Most methods of SLOCC-classification are based on detection or quantification schemes, which signal whether a state is or is not equivalent to another state, but do not present explicitly the SLOCC conversion between equivalent states. The use of sequences of ELOs give the exact steps necessary to convert a state into an equivalent one, a feature that can be important in experimental scenarios, in computational routines or in the construction of examples/counter-examples for demonstrations.        

Physical implementations of ELOs could be relevant from an experimental point of view. Since ILOs are implemented via local filters \cite{localfilters}, ELOs would represent the essential physical processes needed in order to convert a state into another. The number of ELOs used over a SLOCC process could then be used as a quantifier of the number of physical resources needed to perform a task. 

Even though the focus of the article was on analytical results, computational and numerical considerations may play a big role in forthcoming developments. For example, standard GJE is numerically very sensitive to ill-conditioning of the coefficient matrix and its multipartite version might be even more sensitive, due to the repeated application of the procedure. The method is thus probably more powerful for states with symmetries, or coefficients that have clear functional descriptions, as is the case for LME and hypergraph states; random coefficient matrices might be very costly to characterize or even intractable from a computational perspective. These issues open new challenging and interesting questions for future developments on the subject.

\section{Acknowledgments}

This work was supported by FAPEMAT and Governo do Estado de Mato Grosso, Project No. 37730.544.26049.23092016. Special thanks to Nilmara Meirelles for ideas and discussions.   

\section{Appendix - Standard GJE}

A linear system of $m$ equations and $n$ unknowns $x_1,x_2,\ldots,x_n$ can be written as $AX=B$, where $A$ is an $m\times n$ matrix, the coefficient matrix of the system, $X$ is a $n\times 1$ column-matrix containing the unknowns and $B$ is a $m\times 1$ column-matrix. The GJE procedure consists of applying sequences of elementary matrices to the left of $AX=B$, in order to bring the coefficient matrix $A$ to an unique standard form $R(A)$ known as the Row-Reduced Echelon Form of $A$. One possible configuration for a $4\times 6$ coefficient matrix $A$: 
\begin{eqnarray*}
R(A)=\left(
\begin{array}{cccccc}
{0}&{1}&{0}&{*}&{0}&{*}\\
{0}&{0}&{1}&{*}&{0}&{*}\\
{0}&{0}&{0}&{0}&{1}&{*}\\
{0}&{0}&{0}&{0}&{0}&{0}
\end{array}\right)
\end{eqnarray*}
The $*$'s are arbitrary scalars, while $1$'s are the so-called pivots; by elementary row-operations a pivot can eliminate any coefficient that is in the position above it or bellow it, without affecting the other pivots. However, due to the restriction to row-operations only, it is not possible to remove coefficients to the right of a pivot. The columns corresponding to these non-removable coefficients are associated with the free parameters of the system, the unknowns that a new solution for any values they assume. 

From a geometric point of view, systems of linear equations represent intersections of linear subspaces if $B=0$ - one can restrict the discussion to this situation, without loss of generality. Even though the presence of free parameters imply infinite solutions (assuming solutions exist and working in fields of characteristic zero), i.e., each different value of a parameter is a point in the intersection, it is the number of pivots and their distribution in the reduced form $R(A)$ that determines the geometrical situation at hand. 

For example, in $\mathbb{R}^3$ if the coefficient matrix has a Reduced Form with one pivot, the system describes a plane. If there are two pivots, the system describes the intersection of two planes, a straight line. And if there are three pivots, there are no free parameters and we have a unique solution, a point.


\begin{thebibliography}{99}
\bibitem{note}  Neverthless, the action of ILOs/ELOs per se is not enough to completely determine the SLOCC-equivalence of a mixed state, 
since it is a special case of a local filtering operation \cite{multientang} and thus not a general SLOCC map.
\end{thebibliography}
\end{document}